\documentclass[acmsmall,screen=true,review=false, anonymous=false]{acmart}

\AtBeginDocument{%
  }

\setcopyright{acmlicensed}
\copyrightyear{2018}
\acmYear{2018}
\acmDOI{XXXXXXX.XXXXXXX}

\acmConference[Conference acronym 'XX]{Make sure to enter the correct
  conference title from your rights confirmation emai}{June 03--05,
  2018}{Woodstock, NY}
\acmISBN{978-1-4503-XXXX-X/18/06}

\usepackage{pgfplots} 
\pgfplotsset{compat=1.17} 
\usepackage{lscape}
\usepackage{colortbl} 

\usepackage{longtable}
\usepackage{subfigure}
\usepackage{adjustbox}

\begin{document}

\title[How HCI Research  Conceptualizes Accessibility of Virtual Reality in the Context of Disability]{An Equitable Experience? How HCI Research  Conceptualizes Accessibility of Virtual Reality in the Context of Disability}

\author{Kathrin Gerling}
\email{kathrin.gerling@kit.edu}
\affiliation{%
  \institution{Karlsruhe Institute of Technology}
  \city{Karlsruhe}
  \country{Germany}
}

\author{Anna-Lena Meiners}
\email{meiners@kit.edu}
\affiliation{%
  \institution{Karlsruhe Institute of Technology}
  \city{Karlsruhe}
  \country{Germany}
}

\author{Louisa Schumm}
\email{louisa.schumm@student.kit.edu}
\affiliation{%
  \institution{Karlsruhe Institute of Technology}
  \city{Karlsruhe}
  \country{Germany}
}

\author{Jan Rixen}
\email{jan.rixen@kit.edu}
\affiliation{%
  \institution{Karlsruhe Institute of Technology}
  \city{Karlsruhe}
  \country{Germany}
}

\author{Marvin Wolf}
\email{marvin.wolf@kit.edu}
\affiliation{%
  \institution{Karlsruhe Institute of Technology}
  \city{Karlsruhe}
  \country{Germany}
}

\author{Zeynep Yildiz}
\email{zeynep.yildiz@kit.edu}
\affiliation{%
  \institution{Karlsruhe Institute of Technology}
  \city{Karlsruhe}
  \country{Germany}
}

\author{Dmitry Alexandrovsky}
\email{dmitry.alexandrovsky@kit.edu}
\affiliation{%
  \institution{Karlsruhe Institute of Technology}
  \city{Karlsruhe}
  \country{Germany}
}

\author{Merlin Opp}
\email{merlin.opp@student.kit.edu}
\affiliation{%
  \institution{Karlsruhe Institute of Technology}
  \city{Karlsruhe}
  \country{Germany}
}

\renewcommand{\shortauthors}{Gerling et al.}

\begin{abstract}
  Creating accessible Virtual Reality (VR) is an ongoing concern in the Human-Computer Interaction (HCI) research community. However, there is little reflection on how accessibility should be conceptualized in the context of an experiential technology. 
  We address this gap in our work: We first explore how accessibility is currently defined, highlighting a growing recognition of the importance of equitable and enriching experiences. We then carry out a literature study (N=28) to examine how accessibility and its relationship with experience is currently conceptualized in VR research. Our results show that existing work seldom defines accessibility in the context of VR, and that barrier-centric research is prevalent. Likewise, we show that experience -- e.g., that of presence or immersion -- is rarely designed for or evaluated, while participant feedback suggests that it is relevant for disabled users of VR. On this basis, we contribute a working definition of VR accessibility that considers experience a necessary condition for equitable access, and discuss the need for future work to focus on experience in the same way as VR research addressing non-disabled persons does.  
\end{abstract}




\maketitle

\section{Introduction}
Making Virtual Reality (VR) accessible for disabled people is of ongoing concern within the Human-Computer Interaction (HCI) and accessibility research communities. For example, there have been a number of empirical investigations addressing access barriers, e.g., \citet{mott2020} explore whether and how people with limited mobility can engage with VR, showing that the technology is associated with numerous access barriers. This is echoed by \citet{creed2023, creed2024}, who carried out multidisciplinary sandpits with expert stakeholders including disabled people, and identified detailed research opportunities pertaining to VR hardware and software to remove access barriers for disabled people. Likewise, Gerling and Spiel \cite{gerling2021} engaged in a theoretical examination of VR from the perspective of disability studies, highlighting that VR is a technology that places high demands on human bodies, which aligns with previous work reflecting on VR accessibility for different groups of disabled people \cite{mott2019}. Here, \citet{dudley2023} highlight the need for \textit{inclusive immersion} in their recent literature review that surveyed VR and augmented reality research, suggesting that we need to move toward \textit{"maximising the inclusiveness of VR and AR technologies"}. However, while their work provides an extensive overview of existing systems, it does not explore the experiential domain of VR in the context of disability.

This raises the question of what experiences disabled people are currently afforded by VR systems, and how experience is addressed in the context of accessibility research: The vision behind VR is one that deeply prioritizes the experiential qualities of the technology \cite{steuer1992}, for example emphasizing the relevance of presence or \textit{the sense of actually being in the virtual environment} \cite{slater1997} as one of the pillars of VR, and details of the human experience of VR are extensively studied in the context of the medium for non-disabled users (also see section \ref{sec:rw_vr_pillars}). Yet, it remains unclear how the experiential domain of VR is approached in HCI research addressing disabled people, and whether experience plays a role in how accessibility is conceptualized. To address this gap, we raise the following two research questions:

\textbf{RQ1:} How does the HCI and accessibility research community currently conceptualize accessibility of VR for disabled users? 

\textbf{RQ2:} What role does \textit{experiential accessibility} or the opportunity for disabled people to have equitable experiences in VR play?

We address these questions through a two-step research process: We first explore how accessibility is currently defined in HCI research and beyond, highlighting a growing recognition of the importance of equitable and enriching experiences. We then carry
out a literature study (N=28) and engage in Qualitative Content Analysis \cite{zhang2005} to examine how accessibility and its relationship with experience is currently conceptualized in research that addresses VR for disabled persons. 

Our results show that existing work seldom defines accessibility in the context of VR. Overall, research examining the barriers associated with VR is prevalent addressing concerns around safety and human factors, while there is a lesser focus on potential facilitators that could support accessible and meaningful VR experiences for disabled people. Likewise, we show that experience -- e.g., that of presence or immersion (cf. section \ref{sec:rw_vr_pillars}) -- is rarely designed for or evaluated, which is a notable deviation from VR research addressing non-disabled persons in which experience is routinely considered. However, we observed numerous instances of disabled participants discussing the importance of experience without being prompted by researchers, underscoring its relevance for all user groups. 

On the basis of these results, our work makes the following three core contributions: (1) We provide a working definition of VR accessibility that accounts for safety, but considers experience a necessary condition for equitable access. (2) We discuss the experiential domain of VR, critically appraising core assumptions underpinning the technology in the context of disability to arrive at an inclusive perspective on the medium. (3) We present opportunities for future work to focus on experience in the same way as VR research addressing non-disabled persons does, outlining how our community can address previous calls for accessibility research to embrace third-wave HCI.

\section{Background}
\label{sec:relatedlit}
In this section, we give an overview of relevant related work. First, we explore current definitions of accessibility through the lens of societal perspectives, legal frameworks, and the HCI and accessibility research communities. Second, we discuss the vision behind VR as an immersive technology, and we summarize the most common intended experiences of the technology. We conclude with an overview of ongoing conversations addressing the accessibility of immersive media and VR.

\subsection{Defining Accessibility: Societal, Legal, and Research Perspectives}
\label{sec:rw_definitions}
Accessibility is a term widely used to describe whether disabled people have equal opportunity to engage with spaces, objects, or experiences. Here, we discuss its use by societal stakeholders, within legal frameworks, and in the HCI and accessibility research communities.

\subsubsection{Societal Perspectives on Accessibility}
There exist a range of definitions of the term accessibility with unique nuances, and colloquial use - i.e., when people describe something as \textit{accessible} - is often inconsistent. With respect to everyday language, Merriam-Webster defines the term \textit{accessible} as the point at which something can be \textit{"easily used or accessed by people with disabilities"} \cite{merriam-webster2024}, whereas the Cambridge Dictionary explains accessibility as \textit{"the quality of being able to be entered or used by everyone, including people who have a disability"} \cite{merriam-webster2024}. 

Disabled activists often find that mainstream definitions and practices of accessibility fall short of creating truly equal experiences, as accessibility is frequently treated as an afterthought. Disability justice advocate Sarah Jama highlights this by noting, \textit{“When people talk about accessibility, it’s usually around how we build a world around this pre-existing society that fits people with disabilities.”} \cite{lucchetta2019} Yet, from the disability justice perspective, we must build an accessible world that is free and fits everyone \cite{lucchetta2019}. In this context, the disability justice group Sins Invalid \cite{berne2005} introduced the idea of \textit{transformative access}, a concept that redefines accessibility by advocating not only for structural adjustments but also for a shift in societal norms to dismantle barriers and embrace disability as part of human diversity. This approach asserts that accessibility should foster equitable spaces where disabled individuals can actively participate, experience, and influence, and hence take part in shaping society rather than merely accessing available services in a passive role \cite{dancersgroup2023}.

\subsubsection{Legal Perspectives on Accessibility}
\label{sec:rw_def_legal}
Legal frameworks have attempted to provide definitions and explanations of accessibility. For example, the United Nations Convention on the Rights of Persons With Disabilities offers the following explanation under Article 9, Accessibility:
\begin{quote}
    \textit{To enable persons with disabilities to live independently and participate fully in all aspects of life, States Parties shall take appropriate measures to ensure to persons with disabilities access, on an equal basis with others, to the physical environment, to transportation, to information and communications, including information and communications technologies and systems, and to other facilities and services open or provided to the public, both in urban and in rural areas.} \cite[p. 9]{unitednations}
\end{quote}
Explicitly referring to digital products, it further specifies that accessibility includes \textit{"[...] access for persons with disabilities to new information and communications technologies and systems, including the Internet"} \cite[p. 10]{unitednations}. Local legislation translating the convention frequently pick up on core aspects. For example, as part of the EU Strategy for the rights of persons with disabilities 2021-2030 \cite{eu_accessibility_act}, the European Accessibility Act \cite{EUR-Lex} provides a framework for the provision of accessible goods and services that is set to regulate provision of accessible hardware and software products. Specifically addressing the design of user interfaces and system functionality, Annex I highlights that
\begin{quote}
    \textit{[t]he product, including its user interface, shall contain features, elements and functions, that allow persons with disabilities to access, perceive, operate, understand and control the product,}
\end{quote}
thereby implicitly defining accessibility. Likewise, the German \textit{Behindertengleichstellungsgesetz} defines accessibility of \textit{information technology (IT) systems} as the point at which systems
\begin{quote}
   \textit{can be used in a typical way, without particular difficulty, and in principle without the help of others, with the use of assistive technology being permissible. [author's own translation]} 
\end{quote}
For an overview of US legislation on accessibility, please see \cite{mack2021}.

Overall, while the frameworks above have successfully defined areas of relevance in the context of accessibility, they remain vague as to when equitable access is achieved. Here, we observe that it is defined against non-disabled experience (i.e., the UN suggesting that accessibility is achieved when disabled people are provided with access comparable to that of non-disabled people \citep[p.9]{unitednations}), while also deeming a lesser experience acceptable (i.e., the absence of \textit{particular} difficulty in German law).

\subsubsection{Perspectives on Accessibility Within HCI Research}
\label{sec:rw_perspectives}
Definitions of accessibility in the HCI research community widely reflect language and perspectives of legal frameworks. In their in-depth exploration of digital accessibility, \citet{lazar2015} define \textit{accessible IT} as
\begin{quote}
    \textit{[disabled people] having access to the same functions and the same information (not edited or summarized information) at the same time and at the same cost with an ease of use substantially equivalent to that experienced by the general population without disabilities.}
\end{quote}
Along the same lines - albeit less specific - the goal of \textit{digital accessibility} is defined as \textit{"[...] equal access to all kinds of digital systems and services to as many people as possible, including those with disabilities"}, as cited in \cite{inal2020, sharma2020}, a definition which is used in a range of HCI projects addressing digital accessibility (e.g., \cite{goldenthal2021, oswal2019a}). In this context, standardization attempts take a similar direction, although not explicitly including the term disability. Within ISO 9241-11:2018 (ergonomics of human-system interaction), accessibility is defined as the
\begin{quote}
    \textit{extent to which products, systems, services, environments and facilities can be used by people from a population with the widest range of user needs, characteristics and capabilities to achieve identified goals in identified contexts of use},
\end{quote}
where the contexts include \textit{"direct use or use supported by assistive technologies"} \cite{ISO9241-11}.

There have also been efforts to understand accessibility in a broader context. For example, \citet{shinohara2017} presents the concept of \textit{social accessibility}, which seeks to capture social factors that should be taken into account when designing assistive technology, e.g., the impact of the presence of others and the role of stigma when systems are used. Advancing the aspiration behind accessibility efforts, \citet{oswal2019}  explicitly addresses the relevance of user experience (UX), suggesting that for a digital system to be truly accessible, the experience that disabled users can achieve needs to be taken into account. In their work, Oswal provides the example of a screen reader user who can -- in principle -- access textual information on a website but cannot experience many of the additional visual elements we typically find on the web in a meaningful way. Likewise, there have been attempts to integrate considerations regarding accessibility and UX. For example, \citet{sauer2020} suggest the term \textit{interaction experience} as an umbrella concept comprising accessibility, usability, and UX; however, the authors do make an explicit distinction between accessibility and higher-level aspects of UX. Nevertheless, this highlights the relevance of disabled users' subjective quality and extent of experience with digital technology in the context of equitable access to digital technology.

Notably, a significant amount of research addressing accessibility does so without explicitly defining the concept. For example, Mack et al.'s \cite{mack2021} recent literature survey of accessibility research within the HCI community extensively uses the term accessibility, but the authors never provide a clear definition of it. Likewise, other literature studies, e.g., the one by \citet{brule2020}, or a co-word analysis that addresses accessibility research by \citet{sarsenbayeva2023}, center accessibility within their work, but do so without provision of a definition, once more highlighting the need to develop language to comprehensively reflect on accessibility in the context of digital technology. 

Specifically addressing immersive media and VR, there have been some attempts to move in this direction. Addressing game accessibility and the importance of player experience, \citet{power2018} propose to foreground \textit{inclusive experiences}, in which users gain basic access to a system, thereby are enabled to achieve their own goals, which forms a basis on which they experience a game -- or have \textit{"fun or other accessible player experiences (APX)"}. Similarly and highly relevant in the context of our work, \citet{dudley2023} define the concept of \textit{inclusive immersion} leveraging it as a lens in a literature review on accessibility in VR and AR. The authors define the concept as \textit{"maximising the inclusiveness of VR and AR technologies"}, and later indicate that it also refers to \textit{"the pursuit of maximally accessible and enjoyable"} systems. However, while they provide an extensive and helpful review of existing VR and AR systems for disabled users and survey design strategies to improve accessibility, they do not focus on user perspectives and the experiences they have with existing technology, or, whether \textit{inclusive immersion} is in fact achieved.  

\subsection{Understanding VR as an Experiential Technology}
\label{sec:rw_vr_as_experience}
When addressing the accessibility of VR technology and attempting to appreciate how \textit{inclusive immersion} can be understood in terms of the experiences that disabled users make with VR, it is relevant to consider the original vision behind VR, and to explore how VR experiences are approached by HCI research in the absence of disability.

\subsubsection{The Vision Behind Virtual Reality}
\label{sec:rw_vr_vision}
Technologies that allow users to transpose themselves into alternative realities have long been discussed in the arts, pre-dating technological research, as succinctly summarized in \citet{berkman2024}'s review of the history of VR. Most prominently, Weinbaum's novel \textit{Pygmalion's Spectacles} \cite{weinbaum2007} dates back to the mid-30s of the 20th century and evolves around a pair of glasses that allow readers to directly enter into and participate in movie-like stories, which one of the main characters within the work describes as
\begin{quote}
    \textit{a movie that gives one sight and sound. Suppose now I add taste, smell, even touch, if your interest is taken by the story. Suppose I make it so that you are in the story, you speak to the shadows, and the shadows reply, and instead of being on a screen, the story is all about you, and you are in it. Would that be to make real a dream?}
\end{quote}
This excerpt, on the one hand, highlights the vision behind the technology -- a utopian aspiration of \textit{making a dream come true} -- and on the other hand, addresses the experiential qualities associated with it, i.e., sensory immersion and first-person interactivity. Likewise, other works of science-fiction have addressed the idea of immersing oneself in virtual worlds, for example, Gibson's \textit{Neuromancer} \cite{gibson1984}, and Broderick's novel \textit{The Judas Mandala}~\cite{broderick1982}, through which he alledgedly was the first to introduce the term \textit{Virtual Reality}. Similarly, the popular science-fiction series StarTrek \cite{startreck} introduces the idea of a \textit{Holodeck}, a fictional technology that leverages holograms to simulate an environment in which users can move around and make experiences within freely and without the need to wear additional hardware on board of its starships. Here, the simulated environment is intended to create an illusion of reality, and predominantly enables training, and leisure and restoration \cite{holodeck}. Beyond these two prominent examples, pop culture has repeatedly picked up the theme of simulated reality in the context of possible futures, mixing technological developments and original vision. For example, Cline's 2011 novel (later turned movie) \textit{Ready Player One} \cite{ready_player_one} leverages head-mounted virtual reality in combination with haptic suits that resemble today's technological efforts and depicts it as a technology that captivates users, and the dystopian production \textit{Black Mirror} (episode: Striking Vipers \cite{striking_vipers}) depicts a variant of head-mounted VR with real-life consequences (e.g., pain) in which two characters have a sexual encounter.

In parallel to these developments in popular culture, the academic community likewise established visions of Virtual Reality, in some cases closely linked with the above-mentioned examples. Most notably, Sutherland's 1965 vision of the \textit{ultimate display} \cite{sutherland1965} describes a technology that creates a perfect illusion of being in another place, touching upon Weinberg's vision \cite{weinbaum2007}. Along the same lines, the Holodeck and its capability to transpose users into virtual environments was acknowledged as the \textit{essence of VR} and an \textit{unattainable holy grail} by \citet{heim1995}, highlighting the gap between vision and technological reality in the 1990s.  Finally, bridging into psychology and the experience of VR, \citet{biocca1995} contemplate the need for \textit{physical transcendence}, i.e., moving beyond the boundaries of the physical world, and fully transposing the bodies of users into the virtual. At the same time, the vision was put into life by industry stakeholders such as Lanier and Zimmerman's VPL Research, introducing a technical focus on VR, with Lanier later defining the medium as a
\begin{quote}
    \textit{three-dimensional, computer generated environment which can be explored and interacted with by a person. That person becomes part of this virtual world or is immersed within this environment and whilst there, is able to manipulate objects or perform a series of actions.}\footnote{https://www.vrs.org.uk/virtual-reality/what-is-virtual-reality.html}
\end{quote}
In response to a growing body of hardware-centric views guiding the further development of VR, \citet{steuer1992} reiterated the perspective from psychology and communication research, building upon the work of Gibson~\cite{gibson2014}, who explored presence, or \textit{"the sense of being in an environment"} \cite[p. 75]{steuer1992}. On this basis, Steuer argues for a definition through the lens of experience, suggesting that \textit{"A virtual reality is defined as a real or simulated environment in which a perceiver experiences telepresence."} \cite[p. 75-75]{steuer1992}, i.e., a mediated experience of presence in an alternative reality, perhaps making the strongest argument to date to explore the human experience of VR.


Overall, we conclude that the vision behind VR is one that necessitates hardware suitable to create the illusion of being in a virtual world, driven by a desire to enable users to experience virtual worlds and relationships as real, thereby becoming a part of the virtual, and being both physically and emotionally affected by it.

\subsubsection{The Pillars of Experience in VR}
\label{sec:rw_vr_pillars}
Building upon the vision behind VR, the wider HCI research community has engaged in comprehensive efforts to provide technologies capable of transposing users into virtual worlds. In this context, pillars of VR experience have been operationalized, focusing on constructs that allow us to design for and evaluate the transposition of human users into virtual worlds. Most notably, this is related to the concepts of immersion and presence. Because both terms have been used and defined ambiguously and slightly differing in different research domains \cite{nilsson2016, agrawal2019}, it is necessary to note how we understand these concepts.

In VR research, \textbf{immersion} is widely comprehended as the objective level of sensory fidelity a VR system provides \cite{slater2003}, and \textit{being immersed} as a psychological state highly dependent on the technological properties of a system that lead to a user’s perception of being \textit{“enveloped by, included in, and interacting with [the provided] stream of stimuli and experiences”} \cite{witmer1998}.

This \textbf{presence} then refers to \textit{"a user's subjective psychological response to a VR system"}  \cite{bowman2007}, specifically their experience of actually being in the virtual environment and detached from the physical world \cite{slater1997}. \citet{lee2004} defines three types of presence: a "physical", a "social", and a "self presence" -- each referring to which virtual artifacts users experience as actually "being there", meaning: virtual objects and surroundings, social actors, or users themselves can be perceived as physically there. Therefore, immersion can be viewed as one concomitant prerequisite to the perception of presence.

Furthermore, to achieve (self) presence in virtual worlds it is necessary to adequately represent the users within VR \cite{huang2023}. This is typically achieved through the use of avatars, an embodiment of the user in the virtual environment. Avatars oftentimes come as full-body representations, but half-body or hand/arm representations are not uncommon, especially in first-person simulations \cite{weidner2023}. Related to the use of avatars is the question of \textbf{body ownership} or \textit{"the special perceptual status of one's own body"} \cite{tsakiris2010}. Here, body ownership illusion \cite{slater2010} refers to perceiving another body -- or digital representation of a body -- as one's own, which is supported by the exploitation of sensory and psychological phenomena. A popular early example is the Rubber Hand Illusion \cite{botvinick1998}. The role of avatars for the experience of VR has been addressed extensively by the HCI research community, e.g., regarding the perception of self in Social VR \cite{freeman2021}, how changing one's VR avatar may increase one's creativity \cite{derooij2017}, and how "completeness" of a virtual body influences the experience of embodiment \cite{feick2024}. In a few case studies in Social VR, the specific preferences of disabled users when designing VR avatars for themselves were discussed \cite{angerbauer2024, mack2023}. The broad consensus of all of these studies is that users' needs and preferences regarding avatar representation vary heavily depending on context and system of use, and that the avatar design immensely impacts users' experiences with VR.

Beyond these key constructs, HCI research has also explored additional experiential qualities of VR, for example, how different factors influence the experience of exiting VR \cite{knibbe2018}, how VR can be leveraged to let users experience having more-than-human capabilities \cite{sadeghian2021}, or how UX of VR versus real environments relates to the feeling of presence \cite{brade2017}. Here, \citet{kim2020} systematically review current VR research through the lens of UX models and frameworks, showing the need of existing taxonomies and research methods to be extended and refined as VR technology evolves and new interaction techniques and usage contexts emerge.

Overall, this emphasizes the strong focus on experience in ongoing research on VR that does not specifically address disabled individuals, instead focusing on unspecific or non-disabled user groups, demonstrating how even small changes in user representation and VR interaction can have significant implications for the way VR is experienced.

\subsection{A Working Definition of Accessibility in the Context of Virtual Reality}
\label{sec:rw_vr_workingdefinition}
Given the strong experiential dimension of VR outlined above, we now reflect on implications for VR accessibility.

Drawing together the different perspectives on accessibility (see section \ref{sec:rw_definitions}), we conclude that there is consensus that accessibility ensures disabled persons can \textit{somehow} interact with spaces, objects, or technologies on the basic level, but that the \textit{quality} of interaction and the \textit{richness} of experience that users can achieve is only partially accounted for, or considered part of other, distinct constructs such as UX (see section \ref{sec:rw_perspectives}). This is a missed opportunity for interactive technology, which routinely foregrounds the experiences that users can make with it, and where comprehensive access includes the provision of certain experiential qualities (e.g., facilitating the experience of competence or autonomy in the context of digital games \cite{villalobos-zuniga2021}, or enabling the experience of presence and immersion in the case of VR, also see section \ref{sec:rw_vr_pillars}). Here, we align with previous work by \citet{dudley2023, oswal2012, power2018} and \citet{putnam2023} that acknowledges the importance of experience in the context of accessibility. Rather than introducing additional umbrella constructs, we argue that experience should already be incorporated into our basic definitions of VR accessibility. Here, we ask: \textit{How can an interactive technology be considered accessible without taking into account the experience that a user has with it?}

Therefore, we conclude that there exists a gap between the aspirations of VR technology concerning the experiences that it seeks to facilitate for its users, and how we currently conceptualize accessibility in immersive technologies. We thus build upon these existing considerations to derive the following working definition of accessibility in the context of VR: \textbf{Accessibility of VR refers to the absence of barriers that would negatively impact how disabled people interact with and experience Virtual Reality, and is achieved when all user groups have an equitable, high-quality experience that accounts for their abilities}, aligning with the idea of ability-based design \cite{wobbrock2011}. In the remainder of our work, we will build upon our working definition in an exploration of how accessibility is operationalized in existing research into VR for disabled users, with the goal of understanding how VR accessibility is currently approached, and the role that disabled users' experience plays within the HCI and accessibility research communities. 

\section{Literature Study: How does the HCI and accessibility research community approach VR accessibility?}
In this section, we describe how we carried out the literature study to understand how the HCI and accessibility communities currently approaches accessibility in VR research. We first give an overview of how we constructed the literature corpus, and we present our analytical approach. Then, we describe the resulting corpus with respect to publication details, and we include key characteristics such as types of VR systems, target groups, and research approaches to aid the interpretation of results.

\subsection{Corpus Construction}
\begin{figure}[htbp]
\centering
\includegraphics[width=0.9\textwidth]{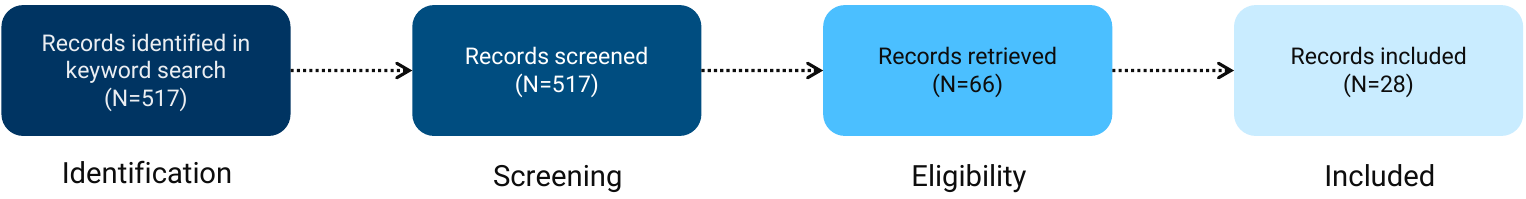}
 \caption{Overview of the PRISMA \cite{page2021} record selection process that we applied.}
\label{fig:prisma}
\Description{Overview of the four-step PRISMA process starting with (1) identification or records identified in keyword search, N=517. (2) screening or records that were screened, N=517. (3) eligibility or records screened, N=66, (4) included records, N=28.}
\end{figure}

Here, we describe how we constructed our corpus for the literature study. We follow the PRISMA reporting guidelines \cite{page2021}, and our process is visualized in figure \ref{fig:prisma}.

\subsubsection{Identification of Relevant Records}
Based on our research questions and engagement with related literature that surveyed accessibility research \cite{dudley2023, mack2021}, we constructed a search query to retrieve relevant research papers.

\textbf{Search Query:} [[Title: "virtual reality"] OR [Title: "vr"] OR [Abstract: "virtual reality"] OR [Abstract: "vr"]] AND [[Title: access* OR disab*] OR [Abstract: access* OR disab*]]

We want to be transparent that we decided against inclusion of specific disabilities in our search query (as for example done in the work of \cite{mack2021}): Descriptions of disability are incredibly broad and we did not want to risk including some while missing others, and given our focus on accessibility and inclusion of it in the search terms, we assumed that we would thereby retrieve relevant papers. Likewise, we limited our search to title and abstracts of papers to only include those in our work that prominently address accessibility. This was also necessary given that accessibility is an overloaded term which is routinely used in papers with no reference to disability.

Aligning with previous literature studies in HCI \cite{spiel2021a, thieme2020}, the search was carried out on May 17th 2024 on the ACM Digital Library Guide to Computing Literature, and yielded an initial 517 results. 

\subsubsection{Screening, Eligibility, and Included Items}
To guide our screening process, we developed a set of inclusion and exclusion criteria in line with our research questions, critically appraised within the author team. 

On this basis, we included papers that addressed immersive virtual reality through original research (IC1), with disabled people being the key audience of the work (IC2). Here, we note that we follow the WHO definition of disability and also include people with chronic illness that is considered disabling \cite{worldhealthorganization2018}. Furthermore, we focused on works that make a contribution to the design of VR (IC3) rather than applying VR to achieve other goals, e.g., optimizing therapeutic outcomes. We excluded those works where accessibility concerns were not in the context of disability (EC1). We also excluded papers focusing on older adults without consideration of disability (EC2); for a detailed appraisal of the difference between disability and old age, please see \cite{knowles2021}. We likewise excluded work focusing on patients, i.e., addressing illness without specific consideration of disability (EC3). We furthermore excluded work that did not focus on immersive VR, e.g., in the medical field, the terminology is sometimes used differently, referring to all interactive systems as virtual reality (EC4), and we excluded works that were not full papers (EC5), or not written in English (EC6).

We applied these criteria to the initial 517 results, screening paper titles and abstracts. The largest share of papers was excluded on the basis of EC1, i.e., we removed works that made reference to access in contexts other than disability, e.g., \textit{access control} in security research. There were no duplicates that we removed. At this point of our process, we retained 66 papers for a full read. In the following stage, we removed papers that did not primarily address disabled people (EC2, EC3), and those not about immersive VR (EC4). We also excluded another thirteen works that were not full papers, but that we could only identify when accessing the full document (EC5). These decisions were discussed within the author team, and we retained 26 records. 

In a final step, we screened the references of included records for further papers that our search may have missed, following a snowballing approach. Here, we identified another two records which -- after discussion within the author team -- were added to our corpus, leading to a final number of 28 records for inclusion in our literature study.

\subsection{Data Analysis}
We analyzed data applying Qualitative Content Analysis following \citet{zhang2005}, which allows us to holistically examine the role that accessibility plays in the existing literature. First, we inductively developed categories in line with the research questions, \textit{RQ1: How does the HCI and accessibility research community currently conceptualize accessibility of VR for disabled users?} and \textit{RQ2: What role does experiential accessibility or the opportunity for disabled people to have equitable experiences in VR play?} This was theoretically underpinned by our exploration of accessibility (see section \ref{sec:relatedlit}). We then applied the categories to five papers from our corpus, discussed the results within the research team, and adjusted the categories, leading to a final set of six categories which were applied to all papers included in our review, a process which was led by the main author of this work. Afterwards, we checked consistency of our codes by revisiting assignment of codes across the entire corpus. To establish trustworthiness \cite{zhang2005, stahl2020}, we discussed the resulting codes and categories within the research team to ensure consensus, which is common practice in a predominantly interpretative research approach. We further provide our coding agenda as proposed by \cite{mayring1994}, which includes categories, their definitions, and examples (see table \ref{tab:coding_agenda}). We also give a detailed overview of our corpus (see appendix \ref{appendix} and supplementary materials) for others to be able to assess our work in more detail. 

\subsection{Positionality}
Considering the qualitative research approach, we want to make explicit our own positionality to allow readers to better interpret our work, being mindful of the challenges associated with this approach \cite{gani2024}. Most importantly, we have previously researched VR for disabled persons, and we have also explored digital games in the context of disability. As such, we believe that meaningful experience -- for example, being challenged, feeling curious, or simply immersed in an interactive environment -- is a relevant design goal. Likewise, our author team includes disabled and non-disabled researchers from different cultures and academic backgrounds (e.g., computer science, psychology, and design), bringing a breadth of perspectives to this research, including experience with the (in)accessibility of VR. 

\begin{table*}[]

\begin{tabular}{p{0.03\textwidth} p{0.2\textwidth} p{0.345\textwidth} p{0.345\textwidth}}
\toprule
& \textbf{Category} & \textbf{Definition}                                                             & \textbf{Examples} \\     
\midrule
RQ1  &  \textit{C1: Definition of accessibility}   & {Definition of the term accessibility, for example on a general level, aligning with previous work (see section \ref{sec:rw_definitions}), or tailored to the context of the specific research.}  & There were no examples in the data. We would have expected definitions along the lines of those presented in section \ref{sec:rw_definitions}.  \\
\midrule
RQ1  &  \textit{C2: Operationalization of accessibility}  & {Explanation how accessibility can be achieved in the given context, for example, as part of research questions guiding the work, as rationale for design decisions, or as outcome measures in user studies.}  & {"We maintained the core implementation of these techniques and augmented them with haptic and auditory cues (e.g., collisions represented with sound and vibrations) to support accessible navigation." [P11, p. 4]}  \\
\midrule
RQ2  &  \textit{C3: Design for experience}   & {Mention of the intended experience when describing design decisions, for example, drawing upon the pillars of VR, i.e., immersion, presence, and body ownership illusion (see section \ref{sec:rw_vr_pillars}), other relevant constructs given a specific context, e.g., player experience.
}  & {"This ensured that we delivered not only high-fidelity information but also a highly personalized and accurate experience tailored to each individual user’s preferences, for better immersion and engagement." [P7, p. 6]}  \\
\midrule
RQ2  &  \textit{C4: Evaluation of experience}   & {Assessment of experience (see previous category for examples) in user studies, for example, using quantitative measures such as questionnaires, or as part of qualitative studies, e.g., in interviews.}  & {Reference to enjoyment and invitations to describe experience through interview questions [P9]; “[...] the main goal of Study 1/2 is to evaluate the performance and user experience [...]” [P17]
}  \\
\midrule
RQ2  &  \textit{C5: User perspectives on experience}   & {Reports of instances relevant to experience that were offered by the research participants, for example, comments on core constructs of VR, also without prompt, or remarks that touch upon their individual experience, e.g., expressing (lack of) enjoyment.
}  & {"P1 remarked that “[VT] is easiest to use, but less immersive than [EE].”" [P7, p. 11]; "Participants discussed the immersion-enhancing potential of spatial audio in VR, as it offers directional sounds and a sense of placement" [P15, p. 8]
}  \\
\midrule
RQ2  &  \textit{C6: Researchers' reflections on experience}   & {Reflection on VR experience by research teams in the discussion of their work, for example, appraisal of the experiences an artifact offered to users, discussion of relevance of experience in the context of accessibility, or acknowledgment of limitations with respect to experience.}  & {"These findings demonstrate the importance and complexity of balancing tradeoffs among the original VR experience, accessibility, and developers’ effort [...]." [P16, p. 11]; "We did not measure if the presented method has an effect on the participants’ immersion in the VR environment." [P19, p. 8]
}  \\
\bottomrule
\end{tabular}
\Description{XXX.}
\caption{Overview of our coding agenda including six categories, aligned with our two research questions.}
\label{tab:coding_agenda}
\end{table*}

\subsection{Corpus Description}
Here, we describe the corpus that provided the foundation for our literature study. We first give an overview of publication dates and venues; then, we will present contribution types, addressed disabilities, research focus and artifacts, and methodology.

\subsubsection{Publication Dates and Venues} 
\label{corpus_sec_publication}
\begin{figure}[htbp]
\centering
\includegraphics[width=0.46\textwidth]{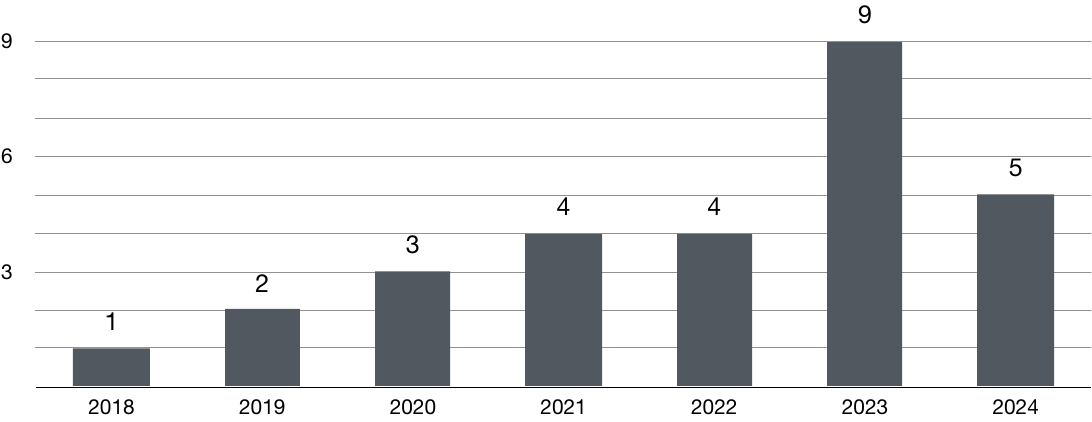}
 \caption{Histogram depicting the publication years for all publications in the corpus.}
\label{fig:corpus_year}
\Description{Histogram using bars showing one publication in 2018, two in 2019, 3 in 2020, 4 in 2021, 4 in 2022, 9 in 2023, and 5 in 2024.}
\end{figure}

Our corpus includes 28 items published over seven years, starting in 2018. While only one publication in 2018 matched our search criteria, an upward trend can be perceived for the following years (see \autoref{fig:corpus_year}). This trend is broken by a stagnation in 2022,  which might be explained by researchers having to adapt their empirical research to the the COVID pandemic. Note that the decline in publications in 2024 originates in the search being executed on the 17th of May 2024 and, therefore, only includes the approximate first third of the year.  

The corpus (see table \ref{tab:corpus}) includes publications from both conferences (21 of 28, 75.0\%) and journals (7, 25\%) with ACM being the main publisher (82\%).  With 9 items (32.1\%), most conference work was published at ASSETS, a conference with a strong focus on accessibility. The preferred journals were TACCESS and TVCG with two publications (7.1\%) each. While TACCESS has a focus on accessibility, TVCG has a broader focus on visualization and computer graphics in general.

\begin{table*}[]
\begin{adjustbox}{width=\columnwidth,center}
\begin{tabular}{llllll}
\toprule
 \textbf{Type} & \textbf{Acronym} & \textbf{Name}                                                             & \textbf{\#} & \textbf{\%} & \textbf{Publisher} \\
\midrule
Conference& ASSETS   & International ACM SIGACCESS Conference on Computers and Accessibility     & 9           & 32,1        & ACM                \\
Conference& CHI      & CHI Conference on Human Factors in Computing System                       & 7           & 25,0        & ACM                \\
Journal& TACCESS  & ACM   Transactions on Accessible Computing                                & 2           & 7,1         & ACM                \\
Journal& TVCG     & IEEE Transactions on Visualization and Computer Graphics                  & 2           & 7,1         & IEEE               \\
Conference& DIS      & ACM Designing Interactive Systems                                         & 1           & 3,6         & ACM                \\
Conference& ICMI     & ACM International Conference on Multimodal Interaction                    & 1           & 3,6         & ACM                \\
Conference& MMVE     & International Workshop on Immersive Mixed and Virtual Environment Systems & 1           & 3,6         & ACM                \\
Conference& PETRA    & PErvasive Technologies Related to Assistive Environments                  & 1           & 3,6         & ACM                \\
Conference& SUI      & ACM Spatial User Interaction                                              & 1           & 3,6         & ACM                \\
Journal& -        & Multimedia Tools and Applications                                         & 1           & 3,6         & Springer           \\
Journal& -        & Universal Access in the Information Society                               & 1           & 3,6         & Springer           \\
Journal& -        & Virtual Reality                                                           & 1           & 3,6         & Springer  \\
\bottomrule
\end{tabular}
\end{adjustbox}
\caption{List of publications indicating publication type, venue, and publisher.\label{tab:corpus}}
\end{table*}

\subsubsection{Contribution Types}
\label{corpus_sec_contrib}
Applying \citeauthor{wobbrock2016}'s \cite{wobbrock2016} taxonomy of contribution types to the corpus (see figure \ref{corpus_fig_contribution}), we found that the majority of the included work made \textit{empirical} contributions (26 of 28, 92,9\%). Most of this \textit{empirical} work also contributed an artifact (20 of 26 \textit{empirical} contributions, 76.9\%). We also found that \textit{artifacts} (20 of 28, 71.4\%) were never the sole contribution, but only occurred in combination with other types of contribution. Through their taxonomy of sound in VR \citet{jain2021a} supplied the only \textit{theoretical} contribution that was nevertheless, again, combined with an \textit{empirical} contribution. While still in the minority, with four papers (14.4\% overall) \textit{survey} contributions were more frequent, for example summarizing specific interaction techniques. Please note that we, as described above, excluded literature studies from the corpus.

\subsubsection{Addressed Disabilities}
\label{corpus_sec_disability}
The work in our corpus focused on different  groups of disabled people. Figure \ref{corpus_fig_disability} gives an overview, with disabilities grouped according to the categories introduced by \citet{mack2021}. In line with their review, we found that people with Motor/Physical impairments (12 of 28, 42.9\%), people who are blind or have low vision (BLV) (10, 35.7\%) and people who are deaf or hard of hearing (DHH) (3, 10.7\%) where the three biggest target groups. Interestingly, while close to each other, our corpus contained more work about Motor/Physical impairment than about blindness and low vision, reversing the findings of \citet{mack2021}. Additionally, \citet{zhang2022} did not further focus on a specific disability but looked into a general representation of disability in VR. Similarly, \cite{creed2023} identified interaction barriers \textit{"across a spectrum of impairments (including physical, cognitive, visual, and auditory disabilities)"} \cite[p.1]{creed2023}. 

\subsubsection{Research Focus}
\label{corpus_sec_focus}
\begin{figure*}[htbp]
\centering
\subfigure[Individual contributions in the corpus divided by types~\cite{wobbrock2011}]{
\includegraphics[width=0.45\textwidth]{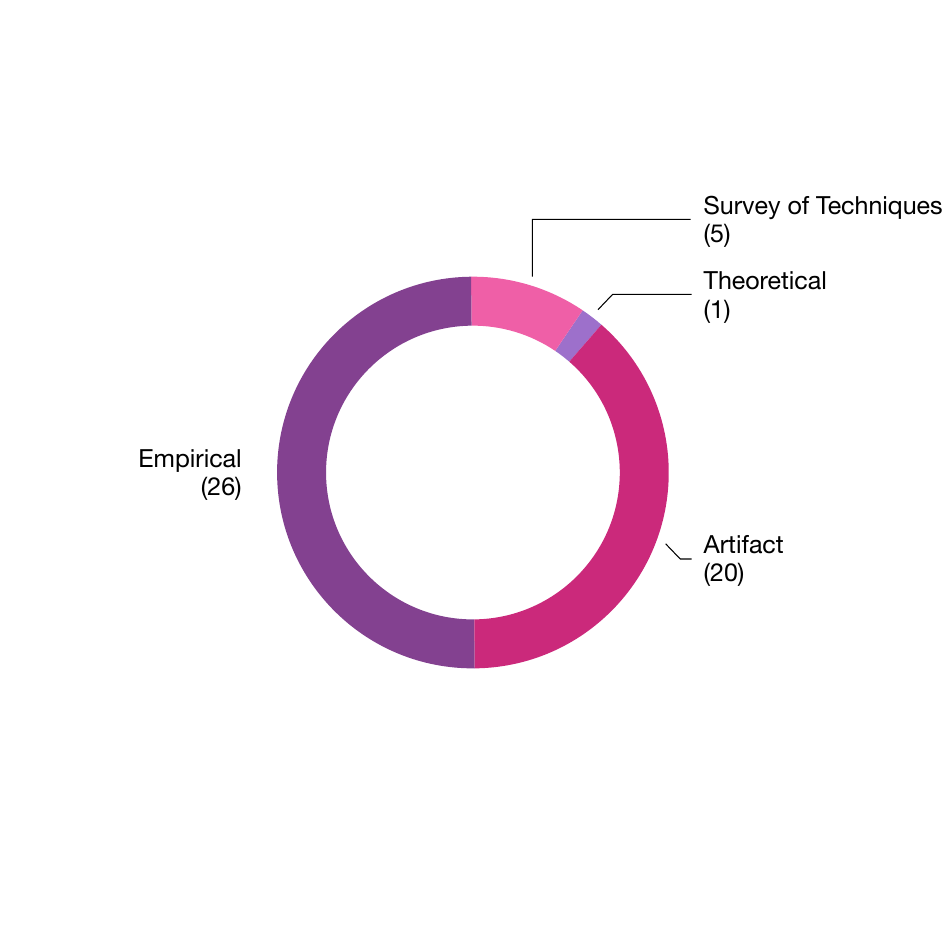}
\label{corpus_fig_contribution}
}
\subfigure[Communities of focus in the corpus (cf.~\cite{mack2021})]{
\includegraphics[width=0.45\textwidth]{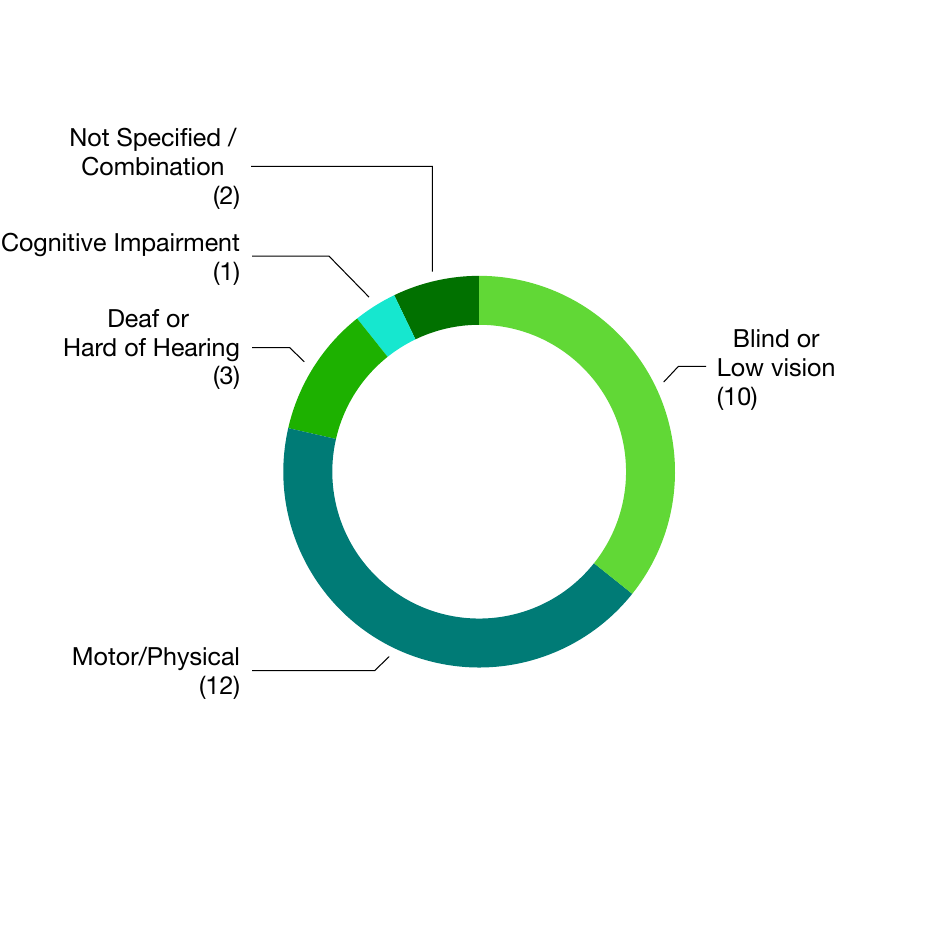}
\label{corpus_fig_disability}
}
\
\subfigure[Focus of research in the corpus]{
\includegraphics[width=0.45\textwidth]{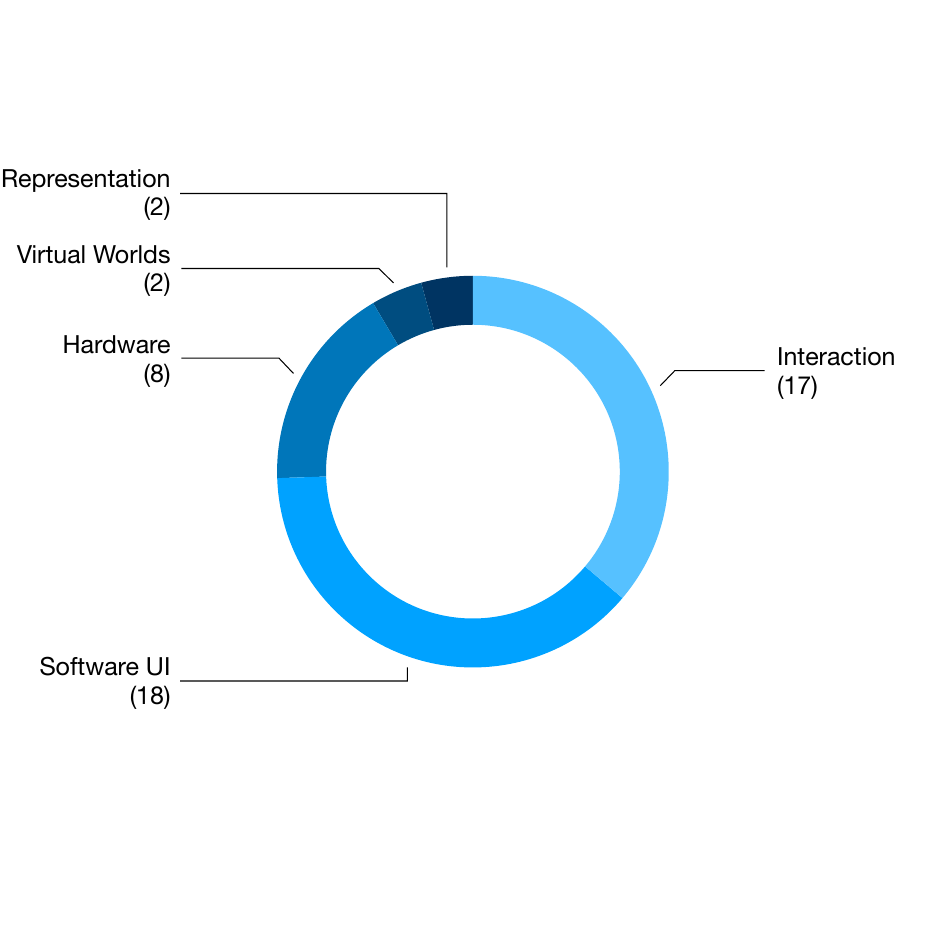}
\label{corpus_fig_focus}
}
\subfigure[Research Methods used in the corpus]{
\includegraphics[width=0.45\textwidth]{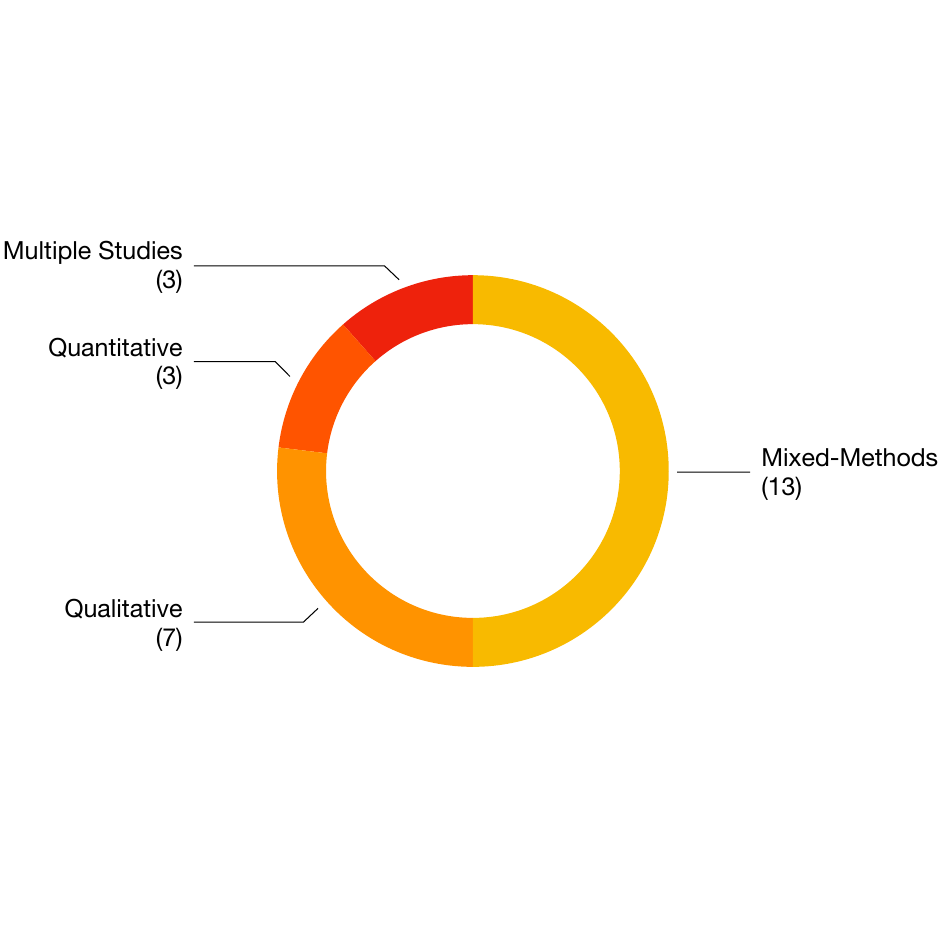}
\label{corpus_fig_methodology}
}
\caption{Characterizing the corpus regarding contribution types, target audiences, research topics, and research methodology. In some cases, papers were assigned to multiple categories (e.g., making an artifact and an empirical contribution).}
\label{fig:main}
\Description{Figure a visualizes contribution types. We observed 26 empirical contributions, 20 artifact contributions, 5 surveys of techniques, and one theoretical contribution. Figure b visualizes communities of focus according to Mack and colleagues. We observed 12 addressing motor/physical disability, 10 addressing people who are blind or have low vision, 3 addressing people who are deaf or hard of hearing, one focusing on cognitive impairment, and two that are unspecified. Figure c addresses the focus of research, with 18 exploring the software UI, 17 interaction paradigms, 8 the hardware, 2 addressing virtual worlds, and 2 focusing on user representation. Figure d finally shows research methods. There were 13 mixed-methods contributions, 7 exclusively qualitative works, 3 exclusively quantitative works, and three works including multiple different studies.}
\end{figure*}

Most of our corpus had either a (partial) focus on \textit{Interaction Paradigms} (17 of 28, 60.7\%), which, among others, contained locomotion (e.g., \cite{weser2023,ribeiro2024}), scene viewing \cite{franz2023}, or general interaction paradigms through upper-body gestures \cite{tian2024}. 17\% (5), in turn, focused solely on \textit{Software UI}, by exploring, e.g., auditory feedback for people who are blind or have low vision \cite{guerreiro2023}, or visual cues to increase balance for people with instable gait \cite{mahmud2023, mahmud2023a}.
All other works had a shared focus on more than one topic, for example, contributing novel VR hardware while also designing new interaction paradigms. The distribution between different focuses is depicted in figure \ref{corpus_fig_contribution}.

While the papers had a specific research focus, most of those were not bound to a specific application context, or such context was not further specified (20 of 28, 71\%). The most named context was \textit{Social VR} (3 of 28, 10.7\%), followed by \textit{VR gaming} (2 of 28, 7.1\%). Further, topics where accessibility assessment of physical spaces \cite{pei2023}, education and entertainment \cite{wu2021}, and one about musical performances \cite{dang2023}.

\subsubsection{VR Hardware}
\label{corpus_sec_artifacts}
All of the work in our corpus used head-mounted VR devices, except two, all in the form of HMDs. While seven works did not further specify which HMD was used, seven stated that they have used the Meta Quest 2 (6) or Oculus GO (1) as stand-alone headsets. In turn, twelve devices were used with external processing units. Here the Vive (5) and Oculus Rift S (3) were used most often. Also, researchers used the Vive Pro Eye, HP Reverb G2 Omnicept, HP Microsoft Mixed Reality VR, and Oculus Rift. The two outliers that did not use HMDs \cite{ji2022,wedoff2019} were focused on the auditory side of VR, instead leveraging specialized headphones to provide spatial audio.

\subsubsection{Methodology}
\label{corpus_sec_methodology}
Of the 26 items that had an empirical contribution (see figure \ref{corpus_fig_methodology}), the most common approach was a mixed-methods study (13, 50.0\%) combining questionnaires to obtain quantitative data with interviews. Qualitative research approaches were likewise prevalent (7, 26.6\%), while only three (11.5\%) publications relied on exclusively quantitative approaches. Our corpus further included works that each contained multiple studies using different approaches, e.g., \citet{gerling2020} first used a qualitative study to gain insights into the motives of wheelchair users to engage with VR, subsequently evaluating a resulting artifact with a mixed-methods study.  

\subsubsection{Study Participants}
While most of the empirical work in our corpus (24 of 26, 92.3\%) recruited participants from the intended target group, thus included disabled persons, two publications included non-disabled people. \citet{jain2021a}, aiming for a taxonomy that supports accessible VR sound representations for deaf and hard of hearing users, arrive at this taxonomy through involvement of hearing sound designers and HCI researchers. For their study about wheelchair locomotion in VR \citet{weser2023} recruited \textit{"fully ambulant participants"} \cite[p.2]{weser2023}. Participants in the included studies were mostly in the range from young adults to middle-aged people. Overall, the participant's gender leaned towards male. Three works (11.5\%) did not report participant gender.

A full overview of the publications included in this review is included in the appendix (see \ref{appendix}).

\section{Results}
In this section, we present the results of the literature study organized along the research questions: First, we address how existing research conceptualizes accessibility in the context of VR and disability. Second, we summarize to which extent the experiential domain of VR is taken into account in HCI accessibility research.

\subsection{RQ1: How does the HCI and accessibility research community currently conceptualize accessibility of VR for disabled users?}
Our results show that the HCI and accessibility research communities only implicitly define accessibility, and that research typically focuses on the removal of access barriers. In the context of empirical work, we observe that this translates into a primary focus on aspects such as user performance and safety.

\subsubsection{Definitions of Accessibility}
Accessibility of VR for disabled users is not explicitly defined in existing work:  \textbf{Out of the 28 publications included in our review, none provided a definition of accessibility, neither more generally nor in the context of VR.} Yet, many works extensively utilize the term throughout their paper. Here, all of the 28 papers made reference to accessibility in the abstract, introduction, or related literature in an effort to motivate their work. For example, we observed general statements about VR accessibility such as by Zhang et al. [P12] pointing out that \textit{"[...] social VR is an emerging but premature medium that lacks sufficient accessibility support [...]" (p.1)}. Closely associated, we observed many instances in which there was a focus on \textbf{barriers and challenges associated with the use of VR for a specific user group}. For example, Yildirim et al. [P6] state that \textit{"Most menu interactions in existing VR applications are designed with the bimanual input assumption in mind and cannot be completed using unimanual input alone." (p. 1)} when addressing limited mobility. Likewise, Mahmud et al. [P19] comment that \textit{"[the obstruction of peripheral vision] is a major accessibility issue for individuals with mobility impairments [...], because VR exacerbates their balance issues, potentially causing falls or injuries." (p. 1)}. The problem-centric research focus is also apparent in the way that research goals are articulated. For example, Mott et al. [P9] point out that \textit{"we must understand the challenges people with limited mobility encounter, or might encounter, when interacting with VR systems" (p. 1)}, while leaving potential instances of people experiencing access unaddressed. 

In contrast to this dominant perspective, \textbf{some authors also appraised facilitators of access under consideration of the opportunities that VR offers}, e.g., Jain et al. [P1] pointing out that sound accessibility is related to \textit{"characteristics such as volume, persistence, and spatial location as well as whether the sound is accompanied by visual or haptic feedback" (p. 2)}. Similarly, Kreimeier et al. [P4] focus on characteristics of VR as an opportunity to create access, pointing out that \textit{"Especially for blind and visually impaired people [the fact that the sensory perception of the environment is computer-simulated] is an [sic] promising possibility to perceive spatial information and overcome limitations of a real objects." (p. 213)}, and Collins et al. [P20] provide an extensive overview of what they term ways of \textit{enhancing accessibility of VR} in the background section of their work on VR accessibility for people who are blind or who have low vision.

\subsubsection{Operationalization of Accessibility}
\label{sec:results_op_accessibility}
\textbf{VR access from a technological perspective in works that made an artifact contribution were most commonly approached with the goal of addressing existing inaccessibilities through design}. For example, Yamagami et al. [P23] articulate (part of) their research contribution as \textit{"[d]evelopment and demonstration of using the creation lens to identify three interaction techniques with the potential to enable accessible control of bimanual interactions" (p. 3)}. Likewise, Ribieiro et al. [P11] explain their design rationale for locomotion techniques for blind people, pointing out that \textit{"We maintained the core implementation of these techniques and augmented them with haptic and auditory cues (e.g., collisions represented with sound and vibrations) to support accessible navigation." (p. 4)}. 

This is also reflected in empirical work, where \textbf{16 out of 20 papers making artifact contributions directly inquired about the accessibility of systems}, for example in the context of accessible VR music performances, where the study and results thereof center around accessibility concerns, e.g., \textit{"Interviewees showed varied levels of VR understanding. Some individuals have seen it on TV or recognized it as a means to experience virtual worlds through glasses or headsets, while others are uncertain about its functioning or have not explored it due to accessibility concerns." [P15, p. 7]} In addition, \textbf{we observe that quantitative measures employed to act as proxies for accessibility typically focused on  usability} (e.g., [P13] and [P17] applied the System Usability Scale \cite{brooke1996}, [P14] explored ease of use, and [P10, p. 63] widely addressed hardware and software usability), which was also reflected in interview guides, e.g., Pei et al. [P7] specifically examined usability, asking \textit{"What do you think of the usability of Embodied Exploration? And could you give concrete reasons?" (p. 9)}. Likewise, we observed enquiries into user performance, for example, Franz et al. [P5] included task performance metrics for locomotion techniques, and Mahmud et al. [P19] measure gait performance. Interestingly, we observed that five publications employed the NASA-TLX~\cite{hart1986} or an adapted version thereof as a measure of task load, reflecting the human factors perspective \cite{kosch2023} in current VR accessibility research. 

Finally, \textbf{safety was a concern related to accessibility in many studies}, e.g., operationalized through simulator sickness [P9, P13], and some research teams such as Kreimeier et al. [P4, p. 215] explicitly exploring \textit{"[participants'] feeling of security"} while generally addressing safety in depth in their work, and South et al. [P3] discussing safety concerns and potential harms of VR. This perspective was also mirrored in qualitative research approaches, where interview questions typically centered on barriers while not giving the same attention to potential facilitators of access. For example, Mott et al. [P9] address VR accessibility in the context of limited mobility, and the video elicitation protocol included in the supplementary material supports a problem-focused research approach primarily interested in barriers. Finally, we want to highlight that relatively few studies made comprehensive inquiries into higher-order constructs of participants' experiences with VR.

\subsection{RQ2: What role does experiential accessibility or the opportunity for disabled people to have equitable experiences in VR play?}
Our results show that many research teams acknowledge the relevance of disabled peoples' experiences in and with VR, but only address the construct of experience superficially in the design and evaluation of accessible VR, while disabled people consider it central to their experience. 

\subsubsection{Design for Experience}
Regarding the design of accessible VR, we note that the experience that disabled people would have with VR was only considered in 8 out of 22 publications that made relevant contributions. \textbf{Those works that did take into account experience often did so briefly when providing rationale for design choices.} For example, Franz and colleagues [P2] explain their choice of scene-viewing techniques from the perspective of realism. Here, they highlight the relationship between presence and realism to justify the benefits of realistic interaction paradigms, pointing out that \textit{"There is evidence to suggest that a relationship between the realism of a VE and presence, the feeling of being physically present in a VE, exists [55]. As a result, many VR interactions aim to mimic real-world interactions." (p. 19)}. The argument of realism is also mirrored in other work, for instance, Ribeiro et al. [P11, p. 6] and Franz et al. [P14] also comment on the benefits of realism. Likewise, work by Jain et al. [P22] and Ji et al. [P26] on VR sound accessibility addresses realism, and all authors do so with specific recommendations. For example, Jain et al. [P22] propose \textit{"Sounds for increasing realism: ambient or objects sounds that increase immersion (e.g., river, vehicles)." (p. 3)}, but also highlight the relevance of sounds for aesthetics, beauty, and to influence the user's affective state. Likewise, Pei et al. [P7] provide rationale for customizable avatar design, pointing out potential benefits for immersion and engagement: \textit{"This ensured that we delivered not only high-fidelity information but also a highly personalized and accurate experience tailored to each individual user’s preferences, for better immersion and engagement." (p. 6)} Engagement is also commented on by Wedloff et al. [P24] in the context of player experiencing, commenting on Flow \cite{sweetser2005} as a design goal when creating VR games. 

\textbf{In those publications that do not explicitly address the experiential dimension, we observe a vague exploration of user preferences and a shallow understanding of experience}. For example, Yamagami et al. [P23] explain that \textit{"We implemented prototypes of the three input techniques for two instances of symmetric out-of-phase interactions that we evaluated with people with limited mobility to learn about user preferences for these tradeoffs." (p. 14)}. Likewise, Wu et al. [P21] provide design rationale for news reading, explaining that \textit{"This would accommodate less technologically capable users, avoid overcrowded menu options, and simplify the user experience to allow quick adjustments to the visual space or switch between different visual settings." (p. 27275)}, mentioning experience, but following up with examples that refer to ease of use.

\subsubsection{Evaluation of Experience}
In the context of user studies and evaluations, 17 out of 26 publications making an empirical contribution touched upon participants' experience, but often did so in an open-ended way: Most commonly, research teams investigated experience through interviews, directly asking participants about the experience that they had had or anticipated with a specific VR system. For example, Collins et al. [P20] report that they \textit{"ended the study with a 30-minute interview, in which [they] asked participants to reflect on their experience and discuss possible improvements" (p. 6)}. While some studies explicitly addressed experience, others only made implicit reference to it. For example, Mott et al. [P9] addressed enjoyment and invited participants to describe their experience as part of interviews, Wedoff et al. [P24] enquired into \textit{participant preferences} and \textit{"whether the participant would want to play again" (p. 9)}, and Tian et al. [P28] assess \textit{overall satisfaction (p. 7)} and \textit{agreement scores (p. 9)}. \textbf{Only few studies (5 out of 26) examined experience through the lens of key constructs underpinning the experiential dimension of VR}, i.e., presence, immersion, or body ownership (see section \ref{sec:rw_vr_pillars}). Notably, Weser et al. [P13] applied the full Igroup Presence Questionnaire (IPQ, \cite{schubert2001, IPQ}), however, we want to point out that this was done in a user study that did not include disabled persons despite a research focus on limited mobility and wheelchair use (see corpus overview in table \ref{tab:corpus_detail}). Zhao et al. [P8] also applied the IPQ, but removed items related to visuals given their research focus on VR accessibility for people who are blind or who have low vision (and the inclusion of disabled people in their user study). Along the same lines, Franz et al. [P5] applied one item of Slater's, Usoh's, and Steed's presence questionnaire~\cite{slater1994}, arguing that they \textit{"only included question \#1 because researchers found that this question elicited the most direct response for presence and had high discriminating power" (p. 6)}. Other work did not rely on questionnaires and instead explored presence in a more open-ended way through interview questions, e.g., Franz et al. [P2].

\subsubsection{Disabled Persons' Perspectives on Experience}
\label{sec:results_subjectivexp}
With respect to the experiences reported by participants, we want to highlight that many of them commented on experiential aspects of their engagement with VR, and that \textbf{experience played an important role in the appraisal of VR systems even when not prompted by research teams}. 

Here, in 11 out of the 20 publications included in our corpus that invited open-ended qualitative feedback, participants did comment on their experience, and \textbf{many participants discussed presence and immersion to either explain their experience or preferences}. For example, Dang et al. [P15] highlight that \textit{"Participants discussed the immersion-enhancing potential of spatial audio in VR, as it offers directional sounds and a sense of placement." (p. 8)}. Guerreiro et al. [P18] report participants' discussion of the \textit{absence} of immersion as a negative factor, \textit{"participants commented that this felt short of a fully immersive experience, which could be augmented by realistic sounds that could either provide useful information – the opponent breathing to convey their location – or just background sound (e.g., the crowd cheering)" (p. 2769)}. Concerns about poor immersion are mirrored in work by Jain et al. [P22], reporting that \textit{"participants (5/11) were skeptical of their interference with the aesthetics of the VR apps, which could diminish immersion. For example, 'I am not sure but this text-pop up [of notification sounds] could take me out of the scene and diminish immersion. [Also], what if there are a lot of sounds and we have a big text box which looks awkward...' (R3)" (p. 9)}. In a similar vein, and although Pei et al. [P7] do not explicitly evaluate presence and immersion, they do report participant responses addressing immersion, e.g., \textit{"P1 remarked that '[VT] is easiest to use, but less immersive than [EE].'" (p. 11)}, and there are some references to body ownership and representation, e.g., \textit{"However, for visibility, P2 remarked that 'Appearance and wheelchair personalization doesn't make a difference to me so long as my height is correct.'" (p. 12)}. Likewise, body ownership and representation is addressed by participants in work by Zhang et al. [P12], pointing out that \textit{"As H-P7 indicated, 'I have [a cochlear implant] on [my avatar] all the time really, just because that’s what I do in real life. I like my avatar to represent me as realistic as possible or as close to [myself], so if I have a cochlear implant I’m not ashamed of it.'"}. Reflecting on their experience more generally, there were many other instances reported by research teams that highlight the relevance of positive experiences in VR, e.g., \textit{"P03 said, 'Just I found, I was focusing more on the buttons than the actual environment itself, I think that’s fun to have control like that, but for this scenario, I feel it takes away from the wonderment of just looking around and enjoying the environment' (P03)" [P2, p. 26]}. Contrasting the generally positive perspective on presence and immersion, we note that Zhao et al. [P8] report instances in which \textbf{sensory immersion was seen as a risk by participants who are blind or who have low vision}, requiring further adaptation of VR, \textit{"As V4 explained, 'I didn't have a good
sense of direction where I was at [in the real world]. I can hear roughly where the wall is at, by the way it blocks off
sound in the real world. I didn't have that in the VR world.'" (p. 9)}. Similarly, South et al. [P3] voice concerns with respect to immersion for people with photosensitivity, highlighting \textit{"participants’ concerns about not being able to quickly break immersion, as well as concerns about being expected to use VR for long periods of time in workplace, training,
or medical scenarios" (p. 9)}.

\subsubsection{Researchers' Reflections on the Relevance of Experience}
While not all research teams designed for or evaluated disabled peoples' VR experiences, a substantial share (10/28) of the works included in our corpus offered reflection on the relevance thereof. 

Most prominently, \textbf{there was discussion of the relevance of disabled peoples' experience of VR in the discussion sections of the respective papers}. Often, research teams re-emphasized importance of the pillars of VR experience (see section \ref{sec:rw_vr_pillars}) or leveraged them to explain findings, also in cases where these were not considered in the design or evaluation. For example, Zhao et al. [P16] explicitly point out that \textit{"These findings demonstrate the importance and complexity of balancing tradeoffs among the original VR experience, accessibility, and developers’ effort when designing accessibility guidelines for VR." (p. 11)}, although the authors had not previously designed for these aspects. Likewise, Zhang et al. [P12] outline that \textit{"Our findings echoed the Embodied Social Presence Theory [58] that the embodied avatars and the shared virtual space and activities can affect user perception and bring them to a higher engagement level, and further expanded this theory by providing evidences from the disability perspective." (p. 12)}, and Franz et al. [P5] comment that \textit{"it seems presence affected the preference for a VR locomotion technique"}, and further elaborate that \textit{"This finding suggests that participants weigh trade-offs in accessibility, user experience,
and enjoyment when determining their preference for a locomotion
technique."}. Adopting a critical stance on their own work, Guerreiro et al. [P18] further discuss participant comments on the shortcomings of their system prototype with respect to immersion (p. 2270), and Franz et al. [P14] highlight that participants \textit{"also identified new ones that we did not consider, including (1) input device, (2) VE aesthetics, and (3) uniqueness to VR" (p. 12)}. Curiously, some authors also made generalizing statements, e.g., \textit{"Incorporating avatars and wheelchairs that accurately represent users significantly enhances the sense of immersion." (p. 14)}, but do so on the basis of single comments from qualitative user studies rather than broader quantitative assessments.

Finally, \textbf{some authors  recommend that experiential qualities of VR for disabled people are explored in future work}. For example, Jain et al. [P1, P22] highlight this opportunity, for example suggesting to answer the question of \textit{"How much is the original experience (e.g., immersion, game challenge) preserved?" (p. 9)}. Interestingly, some authors also acknowledge the lack of insights into experience as a limitation of their work. Here, Mahmud et al. [P19] comment that \textit{"We did not measure if the presented method has an effect on the participants’ immersion in the VR environment." (p. 8)} in the respective section of their work. In a similar vein, other work justifies the lack of focus on experiential aspects such as presence with focus on accessibility and usability, e.g., \textit{"we were mainly interested in the accessibility and usability of the technique, so we did not administer additional questionnaires, such as one for presence"} [P2].

\section{Discussion}
In our work, we examined VR accessibility through a theoretical exploration of the concept of accessibility and core constructs underpinning VR experience, supplemented by a literature study of how VR accessibility is currently conceptualized in research. Here, we discuss our findings, focusing on the need to broaden our definition and operationalization of accessibility in the context of VR. Furthermore, we reflect on the pillars of VR experience in the context of disability, and we provide opportunities and recommendations for HCI accessibiltiy research that addresses VR technology. 

\subsection{Broadening Our Perspectives on VR Accessibility}

Our results suggest that there is no shared definition of accessibility in the context of VR, and that there is no consensus as to how to account for experience in technical and experimental work.

\subsubsection{Moving Beyond Functionalist Research Paradigms}
Existing research into VR accessibility strongly focuses on the accessibility of hardware, interaction paradigms and feedback provision through software interfaces, with core concerns of research teams centering around the safety and usability of such systems (see section \ref{sec:results_op_accessibility}), relying on user performance metrics and constructs such as cognitive load to assess the quality of interaction. Here, we want to make very clear that these aspects are all integral to designing accessible VR: everyone should be able to feel safe and competent when interacting with the technology. However, the experiential qualities of VR often remained unaddressed, with authors considering them an opportunity for future work. Here, we must wonder whether that future will ever arrive: While initial definitions of VR such as in Steuer's work~\cite{steuer1992}, already clearly articulated the experiential dimension, Bannon \cite{bannon1992} argued over 30 years ago that we must move past the \textit{"limited [human factors] view of the people we design for"}, and B\o{}dker \cite{bodker2006} highlighted the need to ensure that \textit{"new elements as experience are included"} in our research in her highly-recognized 2006 paper on third wave HCI, 
accessibility research still needs to move forward. In 2009, \citet{hedvall2009} pointed out the absence of what he calls \textit{"accessibility experience"}, i.e., a focus on how accessibility is experienced by users when interacting with technology. In a similar vein, \citet{power2018} suggest that experience must be considered in the design of interactive technology for disabled people, leveraging the example of games, which -- as an immersive medium -- are closely related to VR. However, many years on, these perspectives are yet to become mainstream in research on VR accessibility. Given the peculiar nature of VR and the relevance of experience, it is therefore important for our field to move beyond functionalist research paradigms, likewise addressing the quality of the experience that disabled persons have when interacting with VR.

\subsubsection{Toward a Holistic Definition of VR Accessibility: The Case for Experience as a Necessary Condition for Accessibility}
On the basis of our literature review, we want to revisit the initial working definition of VR, \textit{"Accessibility of VR refers to the absence of barriers that would negatively impact how disabled people interact with and experience Virtual Reality, and is achieved when all user groups have an equitable, high-quality experience, aligned with their abilities."} (see section \ref{sec:rw_vr_workingdefinition}). Existing work provides further insight into what a high-quality experience that aligns with abilities consists of. First, there is the need to \textbf{ensure user safety as a prerequisite for a high-quality experience}. Second, we found that enquiries into barriers led to problem-centric perspectives, potentially omitting \textbf{potential facilitators of meaningful engagement with VR}, which should also be accounted for. Third, those works that did examine experience did so through the lens of presence, demonstrating that \textbf{accessibility research can and should apply the same measures of experience as our community affords when designing for non-disabled persons}. Here, we want to underscore that if we wish to achieve truly equitable access, we cannot decouple considerations of accessibility from the experiential domain as suggested by previous work (e.g., \citet{dudley2023}'s concept of \textit{inclusive immersion} or \citet{power2018}'s \textit{accessible player experiences}): Particularly for technologies that aspire to facilitate experiences, they can only be considered accessible when disabled people can tap into these. This echoes previous calls for the consideration of accessibility from the very start of system design \cite{gerling2021}, which we want to extend with the requirement to account for experience from project start, rather than considering it a secondary objective relegated to future work. Likewise, this needs to be understood as a call to action for policy makers to include experience in legal frameworks (see section \ref{sec:rw_def_legal}) that seek to define requirements for equitable access.

\subsection{Reframing VR Experience From the Perspective of Disability}
Experience within VR is typically examined through constructs that are derived from prominent visions of VR (see section \ref{sec:rw_vr_as_experience}), e.g., immersion and presence, with the notion that increasing these contributes to a better experience. However, in the context of disability, there is some evidence that this needs to be approached with nuance, e.g., more immersion not being desirable for all user groups (see section \ref{sec:results_subjectivexp}). 
Here, we must wonder whether allowing oneself to be fully immersed in a technology, being in a position to trust the designers of that technology that the resulting experience will be safe and enriching, ultimately is a privilege for those within the narrow scope of bodies that VR is currently designed for \cite{gerling2021}. Notably, issues surrounding privilege have previously been discussed in the HCI research community \cite{linxen2021}, e.g., in the context of the experiences that people of color are afforded in digital games \cite{passmore2018}, and normative underpinnings of tangible and embedded interaction \cite{spiel2021}. They are likewise mirrored in how accessibility research is carried out, with disabled researchers drawing upon their own experiences remaining a minority \cite{spiel2022, spiel2020}.
When discussing VR in the context of disability, our community should therefore critically reflect on the positionality and norms of those who articulate visions for the technology and set research agendas. Collectively, we should be willing to challenge core assumptions, re-negotiating what constitutes meaningful VR experiences for different groups of disabled people, not stopping with system design \cite{gerling2021}, but also extending to relevant theory \cite{spiel2021} that underpins our research.

\subsection{Opportunities and Recommendations for HCI Accessibility Research}
Here, we discuss three practical opportunities for future research wishing to center the experience of disabled users when creating accessible VR derived from our theoretical exploration and literature study. First, we focus on the relevance of accounting for experience in an integrated way. Second, we address how to complement barriers to VR use with facilitators thereof. Third, we close with a reflection on how VR accessibility research can draw upon the waves of HCI to set an agenda for future work. 

\subsubsection{Consistently accounting for experience by treating it as inherent accessibility requirement that needs to be measured} To ensure that one's experience with VR is accounted for when considering the accessibility of the technology, we recommend to include experience as an inherent accessibility requirement, spanning system design (e.g., reflecting on how design choices impact experience) and evaluation (e.g., inclusion of measures of experience such as in the work on VR for people with limited mobility by Franz et al. [5] measuring presence). Here, we want to note the importance of developing accessible measures, with Zhao et al. [18] adapting an existing measure to people with visual impairments, suggesting there is a research need for tools that can be adjusted to different types of disability. Going forward, a stronger empirical focus on experience would enable a discussion of VR access on this basis.

\subsubsection{Addressing not just barriers, but also facilitators of VR accessibility}
Despite the long-standing call for ability-based design \cite{wobbrock2011}, much of the current research on VR accessibility focuses on barriers without simultaneously addressing the strengths of users and investigating facilitators of access. While a problem-centric perspective is intuitive given the current state of XR accessibility (also see \cite{dudley2023}) and one that is persistent in ongoing work, e.g., \cite{collins2024}, it is a missed opportunity to identify aspects that could improve how disabled people experience VR, but that do not directly map onto the removal of barriers. Here, there is an opportunity for exploratory work involving disabled people to take a more balanced perspective reflecting value-neutral models of disability \cite{barnes2016}, specifically examining what the characteristics of VR worth engaging with are.

\subsubsection{Embracing the Third Wave of HCI in accessibility research}
Our final recommendation extends beyond VR research, and mirrors previous calls to embrace third wave HCI \cite{bodker2006} in accessibility research \cite{hedvall2009, power2018}, making room for (lived) experience, meaningful participation \cite{spiel2020}, and acknowledging the importance of human connection \cite{bodker2015} (which is also reflected in the ongoing discourse on interdependence in the context of assistive technology \cite{bennett2018}), aligning accessibility research with the wider ambitions of the field of HCI. This offers perhaps the biggest opportunity for future research: We need to afford the work on interactive and immersive technology for disabled people the same nuance and care as when addressing non-disabled persons, placing the same emphasis on experience as an outcome parameter for system evaluation and quality of our research, ultimately living up to the calls for disability justice in HCI \cite{sum2022} that we claim to aspire to.

\section{Limitations}
There are a few limitations that need to be considered in the context of our research. 
We surveyed ACM Guide to Computing Literature because we were primarily interested in how Human-Computer Interaction and accessibility research understand accessibility in the context of VR. However, other fields have also engaged with VR, e.g., from the perspective of disability studies, or medical research, which may warrant an additional exploration in the future. 
With respect to work that specifically positioned itself as rehabilitative or therapeutic, we made the  decision to exclude such papers in an effort to focus on work that addressed user interactions with VR first, and focused on the design and human-centric evaluation of VR. However, research addressing therapeutic application of VR may also hold implications for the design thereof, and could for example also give insights into longer-term user engagement with VR in future explorations. 
Finally, our analysis focused on standard constructs associated with VR. While this offered us a viable opportunity to examine existing VR research, an open-ended analysis of experiences with VR may have given more emphasis to potentially unique perspectives of disabled people.

\section{Conclusion}
VR is a concept and technology that promises an engaging experience by transposing users into virtual worlds, aspiring to fully immerse their senses, purporting the feeling of the user avatar and virtual environment being real. In our work, we have examined whether this experiential dimension of VR is also considered in the context of accessibility, showing that definitions often fall short of users' experiences, and that core constructs such as presence remain likewise underaddressed in VR research addressing disabled people. Thus, our work is a call to action for the HCI accessibility research community, highlighting the need to move beyond human factors considerations in VR accessibility research, adequately addressing the experiential domain of VR so that we continue to work toward equitable access to the technology for everyone.
\begin{acks}
This work is supported by the European Research Council under grant number 101115807, StG AccessVR.
\end{acks}

\bibliographystyle{ACM-Reference-Format}
\bibliography{IAR_AccessVR}

\appendix

\section{Appendix}
\label{appendix}

In the appendix, we include a full overview of our corpus so that readers can get a comprehensive overview of the papers included in the literature study. The table is included in landscape format on the following pages.

\begin{landscape}
\begin{center}
\begin{longtable}{p{0.025\textwidth} p{0.1\textwidth} p{0.05\textwidth} p{0.25\textwidth} p{0.1\textwidth} 
p{0.4\textwidth} p{0.25\textwidth}}

\caption{Overview of the literature corpus, with disabled community of focus included in line with Mack et al.'s classification of accessibility literature \cite{mack2021}. 
A variant with more detail is included in the supplementary materials.}

\label{tab:corpus_detail}
\\ \hline
\textbf{ID} & \textbf{Authors} & \textbf{Year} & \textbf{Title} & \textbf{Venue} & \textbf{Summary} & \textbf{Community of Focus} \\
\hline
\endfirsthead

\\ \hline
\textbf{ID} & \textbf{Authors} & \textbf{Year} & \textbf{Title} & \textbf{Venue} & \textbf{Summary} & \textbf{Community of Focus} \\
\hline
\endhead

\endfoot

\endlastfoot

P1 & Jain et al. \cite{jain2021a} & 2021 & A Taxonomy of Sounds in Virtual Reality & DIS 2021 & The paper proposes a taxonomy for categorizing VR sounds by source and intent, designed to guide the creation of visual and haptic substitutes for auditory information in VR environments. & Deaf or hard of hearing  \\
\hline
P2 & Franz et al. \cite{franz2024} & 2024 & A Virtual Reality Scene Taxonomy: Identifying and Designing Accessible Scene-Viewing Techniques & TOCHI & This study introduces a taxonomy for VR scenes intended to guide design decisions for suitable, accessible viewing techniques. Its applicability is evaluated in a study with users with limited head mobility. & Motor or physical impairment: limited (head) mobility \\
\hline
P3 & South et al. \cite{south2024} & 2024 & Barriers to Photosensitive Accessibility in Virtual Reality & CHI 2024 & Through an interview study, South et al. identify four types of barriers that people with photosensitive epilepsy face when interacting with VR as well as potential benefits and areas for improvement of VR technology. & Other: Photosensitive epilepsy \\
\hline
P4 & Kreimeier et al. \cite{kreimeier2020} & 2020 & BlindWalkVR: formative insights into blind and visually impaired people's VR locomotion using commercially available approaches & PETRA 2020 & In this study, Kreimeier et al. use an adapted version of the NASA-TLX questionnaire to compare the perceived usability of four different VR input devices for locomotion, e.g. controllers or treadmills, for blind and visually impaired users. & Blind or low-vision \\
\hline
P5 & Franz et al. \cite{franz2023} & 2023 & Comparing Locomotion Techniques in Virtual Reality for People with Upper-Body Motor Impairments & ASSETS 2023 & In this paper, six locomotion techniques are evaluated with users with an upper-body motor impairment in terms of different user experience factors through quantitative and qualitative measures. Design recommendations for accessibility are derived from the results. & Motor or physical impairment: upper body \\
\hline
P6 & Yildirim \cite{yildirim2024} & 2024 & Designing with Two Hands in Mind? A Review of Mainstream VR Applications with Upper-Limb Impairments in Mind & MMSys 2024 & Building upon P23 \cite{yamagami2022}, this study reviews 16 VR applications with varying purposes, such as productivity or collaboration,  for accessibility by users with upper-limb impairments, focusing on the assumption of bimanual input. Findings reveal that over half of the applications require two-hand use, and none provide customizable unimanual input options. & Motor or physical impairment: upper limbs \\
\hline
P7 & Pei et al. \cite{pei2023} & 2023 & Embodied Exploration: Facilitating Remote Accessibility Assessment for Wheelchair Users with Virtual Reality & ASSETS 2023 & The authors introduce "Embodied Exploration," a VR system designed to enable wheelchair users to remotely assess the accessibility of physical environments in terms of visibility, locomotion, and manipulation tasks. The system provides high-fidelity digital replicas and personalized avatars intended to simulate physical visits and evaluate accessibility and is evaluated with wheelchair users. & Motor or physical impairment: limited mobility, wheelchair users \\
\hline
P8 & Zhao et al. \cite{zhao2018} & 2018 & Enabling People with Visual Impairments to Navigate Virtual Reality with a Haptic and Auditory Cane Simulation & CHI 2018 & Zhao et al. describe the development and an initial evaluation of "Canetroller", a haptic and auditory VR controller designed to support navigation in virtual environments for people using a white cane. It incorporates resistance, vibrotactile feedback, and spatial audio to replicate real-world cane interactions and facilitate spatial awareness. & Blind or low-vision \\
\hline
P9 & Mott et al. \cite{mott2020} & 2020 & "I just went into it assuming that I wouldn't be able to have the full experience": Understanding the Accessibility of Virtual Reality for People with Limited Mobility & ASSETS 2020 & This study explores VR accessibility challenges faced by users with limited mobility through interviews with 16 participants, identifying seven barriers, such as controller use and headset setup. It discusses participant-suggested improvements and proposes design strategies to make VR more accessible for this user group. & Motor or physical impairment: limited mobility \\
\hline
P10 & Creed et al. \cite{creed2023} & 2023 & Inclusive AR/VR: accessibility barriers for immersive technologies & Universal Access in the Information Society & This paper presents the results of two "sandpits" (e.g., full-day moderated group discussion sessions) with disabled and non-disabled participants, identifying key barriers of AR and VR technology for users with different types of disabilities, i.e., neurodivergence, cognitive, physical, visual, and auditory impairments, along the categories of software and hardware usability, ethics, and collaboration/interaction. & Motor or physical impairment, cognitive impairment, intellectual or developmental disability, autism, other, blind or low-vision, and/or deaf or hard of hearing \\
\hline
P11 & Ribeiro et al. \cite{ribeiro2024} & 2024 & Investigating Virtual Reality Locomotion Techniques with Blind People & CHI 2024 & In this study, Ribeiro et al. evaluate the UX quality of three haptically and auditorily augmented locomotion techniques in VR for blind users. UX quality and performance are assessed using metrics like completion rate and self-reported fun of use, as well as semi-structured interviews. & Blind or low-vision \\
\hline
P12 & Zhang et al. \cite{zhang2022} & 2022 & "It's Just Part of Me:" Understanding Avatar Diversity and Self-presentation of People with Disabilities in Social Virtual Reality & ASSETS 2022 & Using systematic review of popular Social VR apps and interviews with people from the DHH community and people with visual impairments, Zhang et al. evaluate various aspects of avatar embodiment in VR, such as customizability of avatars or accessibility of avatar creation processes. The findings are discussed to give design recommendations. & General disability or accessibility \\
\hline

P13 & Weser et al. \cite{weser2023} & 2023 & Navigation in Immersive Virtual Reality: A Comparison of 1:1 Walking to 1:1 Wheeling & Virtual Reality & In their study, Weser et al. compare two VR locomotion techniques, 1:1 walking and the analog use of a wheelchair, 1:1 wheeling, and find no statistically significant differences in different VR-UX aspects, such as positive/negative affect, simulator sickness, usability, and presence. & Motor or physical impairment: limited mobility, wheelchair users \\
\hline
P14 & L. Franz et al. \cite{l.franz2021} & 2021 & Nearmi: A Framework for Designing Point of Interest Techniques for VR Users with Limited Mobility & ASSETS 2021  & In a video elicitation study, Franz et al. gather user feedback on their prototype, "Nearmi". This framework is designed to support the creation of accessible point-of-interest navigation techniques for users with limited mobility in VR, allowing for alternative interaction methods that reduce the need for head and body movement. & Motor or physical impairment: limited mobility \\
\hline
P15 & Dang et al. \cite{dang2023} & 2023 & Opportunities for Accessible Virtual Reality Design for Immersive Musical Performances for Blind and Low-Vision People & SUI 2023 & This study investigates design opportunities for making immersive musical performances accessible to blind and low-vision users. Using a mixed-methods approach (survey and interviews), it explores users' needs and preferences for VR music experiences, identifying multimodal feedback  and customization as key design considerations. & Blind or low-vision \\
\hline
P16 & Zhao et al. \cite{zhao2019} & 2019 & SeeingVR: A Set of Tools to Make Virtual Reality More Accessible to People with Low Vision & CHI 2019 & Yuhang et al. present "SeeingVR", a set of 14 tools aimed at enhancing VR accessibility for users with low vision, providing visual and auditory augmentations to support scene interaction. Evaluations with low-vision users and VR developers show improvements in task completion speed and accuracy. & Blind or low-vision \\
\hline
P17 & Li et al. \cite{li2022} & 2022 & SoundVizVR: Sound Indicators for Accessible Sounds in Virtual Reality for Deaf or Hard-of-Hearing Users & ASSETS 2022 & Li et al. examine the needs and preferences of DHH users regarding the visual augmentation of sounds in VR. The developed system "SoundVizVR" is designed to assist users in the location and identification of sounds. & Deaf or hard of hearing \\
\hline
P18 & Guerreiro et al. \cite{guerreiro2023} & 2023 & The Design Space of the Auditory Representation of Objects and Their Behaviours in Virtual Reality for Blind People & IEEE Transactions on Visualization and Computer Graphics 2023 & Guerreiro et al. define a design space for auditory representations of objects and their behaviors in VR with the goal to improve accessibility for blind users. They classify auditory cues using 9 categories, creating a framework that guides VR developers in making design decisions. A concurrent user study provides insights into user preferences and challenges in using auditory feedback. & Blind or low-vision \\
\hline
P19 & Mahmud et al. \cite{mahmud2023} & 2023 & The Eyes Have It: Visual Feedback Methods to Make Walking in Immersive Virtual Reality More Accessible for People With Mobility Impairments While Utilizing Head-Mounted Displays & ASSETS 2023 & This study investigates visual feedback techniques to improve balance and gait for VR users with mobility impairments using head-mounted displays. In a user study, metrics like walking velocity, and step and stride length are compared to give design recommendations for visualizations. & Motor or physical impairment: limited mobility, instable gait, Multiple Sclerosis \\
\hline
P20 & Collins et al. \cite{collins2023} & 2023 & "The Guide Has Your Back": Exploring How Sighted Guides Can Enhance Accessibility in Social Virtual Reality for Blind and Low Vision People & ASSETS 2023 & In this study, Collins et al. explore how the use of sighted guides for blind or visually impaired people can be transposed into VR with the goal of making Social VR apps more accessible. The framework derived from physical sighted guides is assessed in a prototypical application. & Blind or low-vision \\
\hline
P21 & Wu et al. \cite{wu2021} & 2021 & Towards accessible news reading design in virtual reality for low vision & Multimedia Tools and Applications 2021 & In this position paper, Wu et al. propose guidelines for accessibility features aimed at improving reading experiences in VR for blind and low-vision users. For that, existing tools are reviewed and a toolbox implementing the recommended features is developed. & Blind or low-vision \\
\hline
P22 & Jain et al. \cite{jain2021} & 2021 & Towards Sound Accessibility in Virtual Reality & ICMI 2021 & Building upon their work from P1 \cite{jain_taxonomy_2021}, Jain et al. develop a design space for multimodal substitutes for sound in VR and preliminary assess its applicability with six visual and haptic VR protoypes for sound accessibility for d/Deaf or hard of hearing users. & Deaf or hard of hearing \\
\hline
P23 & Yamagami et al. \cite{yamagami2022} & 2022 & Two-In-One: A Design Space for Mapping Unimanual Input into Bimanual Interactions in VR for Users with Limited Movement & TACCESS & This study presents the "Two-in-One" design space, which maps unimanual input to bimanual interactions in VR. It categorizes interactions by coordination and computer assistance needs, supporting developers to create interaction techniques that leverage one-handed input to improve accessibility for people with limited mobility. & Motor or physical impairment: limited mobility \\
\hline
P24 & Wedoff et al. \cite{wedoff2019} & 2019 & Virtual Showdown: An Accessible Virtual Reality Game with Scaffolds for Youth with Visual Impairments & CHI 2019 & Wedoff et al. introduce "Virtual Showdown", a VR game designed to be accessible to visually impaired youth by using 3D audio and haptic feedback as primary cues to employ verbal and vibration-based scaffolds. The game is evaluated empirically with the intended users with regards to different measures such as performance and experience quality. & Blind or low-vision \\
\hline
P25 & Mahmud et al. \cite{mahmud2023} & 2023 & Visual Cues for a Steadier You: Visual Feedback Methods Improved Standing Balance in Virtual Reality for People with Balance Impairments & IEEE Transactions on Visualization and Computer Graphics 2023 & In a similar study to their work presented in P19 \cite{mahmud_eyes_2023}, Mahmud et al. develop and evaluate different visual feedback techniques to support stable standing in VR for users with balance impairments. A user study finds preferences for specific types of visual feedback compared to others. & Motor or physical impairment: limited mobility, balance impairment, Multiple Sclerosis \\
\hline
P26 & Ji et al. \cite{ji2022} & 2022 & VRBubble: Enhancing Peripheral Awareness of Avatars for People with Visual Impairments in Social Virtual Reality & ASSETS 2022 & This study presents "VRBubble", an audio-based VR feature aimed to enhance peripheral awareness of Social VR avatars for people with visual impairments. The spatial audio feedback the system provides is based on "social distances", i.e. space around one's avatar that is classified as, e.g., intimately close. It is evaluated against a standard audio beacon feature. & Blind or low-vision \\
\hline
P27 & Gerling et al. \cite{gerling2020} & 2020 & Virtual Reality Games for People Using Wheelchairs & CHI 2020 & Gerling et al. explore challenges and opportunities of VR gaming for wheelchair users, including findings from a survey, the design and evaluation of three VR game prototypes, and implications for the design of VR games with (full-body) interactions. & Motor or physical impairment: limited mobility, wheelchair users \\
\hline
P28 & Tian et al. \cite{tian2024} & 2024 & Designing Upper-Body Gesture Interaction with and for People with Spinal Muscular Atrophy in VR & CHI 2024 & This paper describes an elicitation study in which 12 people with Spinal Muscular Atrophy designed upper-body gestures for 26 common VR commands, with the goal of identifying user-defined gestures and the mental models of people with SMA when designing VR gestures. & Motor or physical impairment: limited mobility, Spinal Muscular Atrophy \\
\hline
\end{longtable}
\end{center}
\end{landscape}
\section{Author Statement}
The present paper is most closely related to the previous CHI 2021 publication of the main author that explored VR from the perspective of disability studies to uncover the assumptions that the technology makes about our bodies \cite{gerling2021}. We believe that the work submitted here is clearly distinct from that piece of research, adopting a different research approach and focus, contributing to our understanding of VR in the context of disability from another perspective. It has not previously been submitted to any other venues. 
\end{document}